\documentclass[reprint,aip,jcp,showpacs,superscriptaddress,floatfix,letterpaper]{revtex4-1}
\usepackage{amsmath,amssymb,amsbsy,amsfonts,amsthm}
\usepackage{bm}
\usepackage{color}
\usepackage{units}
\usepackage{graphicx}
\setkeys{Gin}{width=\linewidth,totalheight=\textheight,keepaspectratio}
\usepackage{tikz}
\usetikzlibrary{calc,arrows}

\def\date{}

\begin{document}

\title{Order and disorder in irreversible decay processes}

\author{Jonathan~W.~Nichols}
\author{Shane~W.~Flynn}
\author{Jason~R.~Green}
\email[]{jason.green@umb.edu}
\affiliation{Department of Chemistry,\
  University of Massachusetts Boston,\
  Boston, MA 02125
}






\begin{abstract}

Dynamical disorder motivates fluctuating rate coefficients in phenomenological,
mass-action rate equations. The reaction order in these rate equations is the
fixed exponent controlling the dependence of the rate on the number of species.
Here we clarify the relationship between these notions of (dis)order in
irreversible decay, $n\,A\to B$, $n=1,2,3,\ldots$, by extending a theoretical
measure of fluctuations in the rate coefficient. The measure,
$\mathcal{J}_n-\mathcal{L}_n^2\geq 0$, is the magnitude of the inequality
between $\mathcal{J}_n$, the time-integrated square of the rate coefficient
multiplied by the time interval of interest, and $\mathcal{L}_n^2$, the square
of the time-integrated rate coefficient. Applying the inequality to empirical
models for non-exponential relaxation, we demonstrate that it quantifies the
cumulative deviation in a rate coefficient from a constant, and so the degree
of dynamical disorder. The equality is a bound satisfied by traditional
kinetics where a single rate constant is sufficient. For these models, we show
how increasing the reaction order can increase or decrease dynamical disorder
and how, in either case, the inequality $\mathcal{J}_n-\mathcal{L}_n^2\geq 0$
can indicate the ability to deduce the reaction order in dynamically disordered
kinetics.

\end{abstract}


\maketitle

\section{Introduction}

Observing the natural world, one constant seems to be that systems have the
capacity for chemical and physical change, though the mechanisms for change
vary. Measuring rates of change is a common approach for deducing the mechanism
of kinetic processes. Complex chemical reactions are an example where the rate
equation governing molecular transformations can be evidence of a particular
mechanism.~\cite{Houston06} For this reason, chemical kinetics provides a set
of methods to quantify the parameters of the rate equation: the rate
coefficient and the order(s).~\cite{SteinfeldFH89} These empirical parameters
define the dynamical equations, which are the typical mass-action rate
equations, provided the reaction system is homogeneous -- as in a solution
phase chemical reaction under standard conditions with uniform concentration(s)
throughout. Rate processes with ``disorder'', however, deviate from traditional
rate equations.~\cite{Ross08} In disordered rate processes,~\cite{Zwanzig90} it
is necessary to replace the rate constant in the rate equation with an
effective time-dependent rate coefficient. The origin of this deviation from
classical kinetics is often microscopic structural and energetic changes in the
host environment, which occur on a time scale comparable to the chemical
change.~\cite{Plonka01}

An early example of disordered or dispersed kinetics is the rebinding of carbon
monoxide to hemeproteins following photodissociation.~\cite{AustinBEFG75}
Experiments showed the complexity of this process is due to a multitude of
energy barriers and reaction pathways. More recent advances in single-molecule
techniques are bringing attention to the disordered decay~\cite{ChakrabartiB03}
of structural correlations in biomolecules during electron transfer
reactions~\cite{LuoAXK06} and enzyme catalyzed
reactions.~\cite{TerentyevaERKHLB12,EnglishMOLLSCKX05,MinELCKX05,FlomenbomVLCEHRNVdK05}
Experimental techniques have also inspired new theoretical approaches to
disordered kinetics~\cite{WangW94,WangW93,WangW95} in the forced
unfolding~\cite{Kuo22062010,ChatterjeeC11} and spontaneous
folding~\cite{LeeSW03,ZhouZSW03} of biomolecules. Disordered kinetics is also
of interest in the charge transfer through molecular assemblies because of
their potential use in molecular electronics, solar cells, and artificial
photosynthesis.~\cite{BerlinGSR08,*GrozemaBSR10} These diverse applications
motivate the search for organizing principles and theoretical frameworks for
extracting mechanistic information that extend chemical kinetics. The present
work develops such a framework.

In the examples above, ``disorder'' can be the result of an intrinsically
time-dependent reaction environment (pure dynamical disorder, homogeneous), a
spatially-dependent reaction environment with an underlying, unknown
distribution of rate coefficients (static disorder, heterogeneous), or both. In
any case, the overall kinetics is describable by a single time-dependent rate
coefficient, with underlying rate coefficients that can fluctuate in time or
space. The reaction order of the overall kinetics, however, is a constant; as
in traditional kinetics, the reaction ``order'' in the rate equation is the
power of the concentration dependence of the rate. While the reaction order can
be evidence of a particular mechanism in traditional kinetics, it is not clear
whether the reaction order might provide mechanistic clues in kinetics with
dynamical disorder. As a step in this direction, our interest here is in the
temporal variation of an effective, overall rate coefficient for disordered
kinetic processes, and how this disorder relates to reaction order.

Relaxation or irreversible decay phenomena have been the primary focus of
theoretical and computational works in disordered kinetics.~\cite{Ross08} There
has been a particular theoretical emphasis on first-order
decay~\cite{Zwanzig90}; however, under some circumstances, disordered processes
can have higher-order rate equations; for example, this was shown for a case of
dispersive, early stage protein folding kinetics.~\cite{Metzler98,Plonka00}
Complicating the mechanistic insights that can come from rates, is that
experimental data may agree with both first- and second-order rate equations
when the rate coefficient is time dependent. A further complication is that
multiple microscopic models can lead to the same macroscopic relaxation
behavior; an example for first-order decay are the mechanisms consistent with
the stretched-(non)-exponential decay.~\cite{KlafterS86} These findings
stimulate our investigation into how meaningful the traditional notion of
reaction-order is for phenomenological, non-exponential kinetic data.

Here we consider the dynamic disorder in the macroscopic kinetics of the
irreversible processes
\begin{equation}
  n\,A \to B\quad\quad\textrm{for}\quad n = 1,2,3,\ldots.
\end{equation}
We assume the concentration, $C_n(t)$, of the species $A$ obey the (non-)linear
differential rate equations
\begin{equation}
  \label{eqn:disorderedratelaw}
  \frac{dC_n(t)}{dt} = -k_n(t)\left[C_n(t)\right]^n
\end{equation}
with time-dependent, effective rate coefficients $k_n(t)$. That is, the rates
are $n^\textrm{th}$-order in $A$. [The subscript $n$ will distinguish variables
and parameters of rate processes of different reaction order.] Since these rate
equations are solvable for any $n$, we will consider the entire class of
reactions. However, the highest known order of a chemical reaction is third
order in traditional kinetics, $k(t)\to k$.~\cite{SteinfeldFH89} In disordered
kinetics, second-order seems to be the highest known order in agreement with
experimental results. For example, power law decays describing recombination
reactions and peptide folding satisfy second-order rate
equations,~\cite{PlonkaKB88,PlonkaBC89,Hamill81} though stretched exponential
functions also fit the available data.~\cite{Metzler98,Plonka00} Generalized
functions, encompassing both stretched exponential and power law decay are also
known.~\cite{BrouersSC06}

\section{Theoretical background}

Static and dynamic disorder lead to an observed rate coefficient, $k_n(t)$,
that depends on time in macroscopic rate equations. The time dependence can
stem from fluctuations in the rate coefficient, fluctuations due to the
heterogeneity of the system and/or the reaction environments.  Measuring the
dispersion of the rate coefficients is therefore a natural approach to quantify
the involvement of the environment in the mechanism or the structural disorder
of the medium.~\cite{Dewey92} The degree of variation in the observed rate
coefficient also quantifies the fidelity of a rate coefficient or rate
equation, assuming rate coefficients that vary less over the observational time
scale are more desirable.

A cumulative measure of disorder that we extend here, originally in
Reference~\onlinecite{FlynnZG14}, is the magnitude of the inequality
\begin{equation}
  \mathcal{J}_n(\Delta{t}) \geq \mathcal{L}_n(\Delta{t})^2
\end{equation}
between the statistical length (squared)
\begin{equation}
  \mathcal{L}_n(\Delta{t})^2 \equiv \left[\int_{t_i}^{t_f}k_n(t)dt\right]^2
\end{equation}
and the divergence
\begin{equation}
  \mathcal{J}_n(\Delta{t}) \equiv \Delta{t}\int_{t_i}^{t_f}k_n(t)^{2}dt
\end{equation}
over a time interval $\Delta t = t_f - t_i$. From these definitions,
$\mathcal{L}_n$ and $\mathcal{J}_n$ are functions of a possibly time-dependent
rate coefficient. The statistical length, $\mathcal{L}_n$, we use is the
cumulative time-dependent rate coefficient over a period of time $\Delta{t}$,
and the divergence, $\mathcal{J}_n$, is the cumulative square of the rate
coefficient, multiplied by the time interval. These definitions are independent
of the particular system and could be applied to any disordered kinetic process
with a time-dependent rate coefficient. The difference,
$\mathcal{J}_n-\mathcal{L}_n^2$, quantifies both the variation of $k_n(t)$ in
time and the fraction of the observation time where those changes occur. As a
measure of disorder, the inequality is complementary to other information
theoretic measures of dynamic disorder.~\cite{LiK13,Chekmarev08}

The representation of $\mathcal{J}_n$ and $\mathcal{L}_n$ we define above is
just one from our earlier work on first-order ($n=1$) rate
processes,~\cite{FlynnZG14} which established a connection between the rate
coefficient and a modified Fisher information.~\cite{Frieden04} For first-order
rate processes, the inequality above has two important features: (1)
$\mathcal{J}_1-\mathcal{L}^2_1$ measures the variation of the rate coefficient
in statically or dynamically disordered decay kinetics over a time interval of
interest, and (2) the lower bound, $\mathcal{J}_1=\mathcal{L}^2_1$, holds only
when the effective rate coefficient is constant. These features are sensitive
to the definition of the time-dependent rate coefficient, $k_n(t)$, but we will
show that defining it appropriately, the inequality applies to irreversible
decay kinetics with non-linear rate laws (i.e., kinetics with total
reaction-order greater than unity). We illustrate this framework with
proof-of-principle models for irreversible decay phenomena of total
reaction-order greater than one.

In macroscopic chemical kinetics, experimental data are typically a
concentration profile corresponding to the integrated rate equation. For a
population of species irreversibly decaying over time, we define the survival
function for $n^\textrm{th}$-order decay,
\begin{equation}
  S_n(t) = \frac{C_n(t)}{C_n(0)}.
  \label{eqn:survival}
\end{equation}
For truly irreversible decay, this is the probability the initial population
survives up to a time $t$. In traditional first-order kinetics the survival
probability, $e^{-\omega_1 t}$, is characterized by a rate constant $\omega_1$.
Linear fitting methods of survival data are a standard approach to find an
empirical rate equation. For example, for a first-order process, a plot of $\ln
S_1(t)$ versus time $t$ is linear with a slope of $-\omega_1$ only if the
traditional first-order rate equation is valid. If the graph is non-linear, it
is natural to search for a higher-order rate equation that gives a linear
graph.~\cite{Houston06} Another approach is to introduce a time-dependent rate
coefficient $k_1(t)$ describing the non-exponential decay, when one suspects
members of the population decay in different structural or energetic
environments or in a local environment that fluctuates in time. This approach
should extend to higher reaction orders, but seems to have received less
attention than the case of first-order kinetics.

We define the effective rate coefficient, $k_n(t)$, through an appropriate time
derivative of the survival function, depending on the reaction-order, $n$,
\begin{equation}
  k_n(t) \equiv
  \begin{cases}
    \displaystyle -\frac{d}{dt}\ln S_1(t)          & \text{if } n = 1 \\[10pt]
    \displaystyle +\frac{1}{n-1}\frac{d}{dt}\frac{1}{S_n(t)^{n-1}} & \text{if } n \geq 2.
  \end{cases}
\end{equation}
In any case, $k_n(t)$ has dimensions of inverse time. The survival function is
dimensionless, which prevents $k_1(t)$ from having a logarithmic argument with
units. While the survival function is not necessary on the grounds of units for
$n\geq 2$, we use it to keep the interpretation of $S(t)$ as a survival
probability and to maintain consistency with our previous
work.~\cite{FlynnZG14}

With these $k_n(t)$, we can express the statistical length, $\mathcal{L}_n$,
and the divergence, $\mathcal{J}_n$, for any reaction order. It is then
straightforward to show these $k_n(t)$, give the bound
$\mathcal{J}_n=\mathcal{L}_n^2$ if $k_n(t)$ is independent of time. As an
example, consider an $n^{th}$-order reaction where $n\geq 2$. The traditional
integrated rate equation is the time dependence of the concentration, $C_n(t)$,
\begin{equation}
  \frac{1}{C_n(t)^{n-1}} = \frac{1}{C_n(0)^{n-1}}+(n-1)\omega_n t.
\end{equation}
When $n\neq 1$, the rate constant, $\omega_n$, has dimensions of
[time]$^{-1}$\,$\cdot$ [concentration]$^{-(n-1)}$. From the general definition
in Equation~\ref{eqn:survival}, the survival function is
\begin{equation}
  S_n(t) = \sqrt[n-1]{\frac{1}{1+(n-1)\omega_n C_n(0)^{n-1}t}} 
\end{equation}
and the definition of the effective rate coefficient is
$k_n(t)\to\omega_nC_n(0)^{n-1}$. This time-independent rate coefficient gives
$\left[\omega_n\Delta tC_n(0)^{n-1}\right]^2$ for both the statistical length
squared and the divergence. That is, the bound $\mathcal{J}_n=\mathcal{L}_n^2$
holds when there is no static or dynamic disorder, and a single rate
coefficient, $\omega_n$, is sufficient to characterize irreversible decay. When
the kinetics are statically or dynamically disordered, the rate coefficient
governing irreversible decay is not constant and one must work with the above
definitions of $k_n(t)$.

\section{Results and discussion}

\subsection{Kinetic model with disorder}

To study the effect of dynamical disorder on kinetics with reaction orders
greater than one, we adapt the Kohlrausch-Williams-Watts (KWW)
model.~\cite{Kohlrausch54,WilliamsW70,Plonka01} The procedure is similar to
that of first-order kinetics where exponential decay is ``stretched'' to
characterize non-exponential decays with a two-parameter survival function
$\exp[-(\omega_1 t)^\beta]$. The parameter $\omega_1$ is a characteristic rate
constant or inverse time scale and the parameter $\beta$ is a measure of the
degree of stretching; both parameters depend on the system and can depend on
external variables such as temperature and pressure.~\cite{Plonka01} This
procedure for modifying traditional kinetics introduces a time-dependent rate
coefficient, $k_{1,\beta}(t) = \beta(\omega_1 t)^{\beta}/t$, into the empirical
model. While different mechanisms can lead to stretched exponential
decay,~\cite{KlafterS86} and any model with a time-dependent rate coefficient
can be subject to our analysis, this type of decay is a good example for our
purposes, given its generality.~\cite{BrouersSC06}

\begin{figure}[b]
  \centering
  \includegraphics[width=0.98\columnwidth]{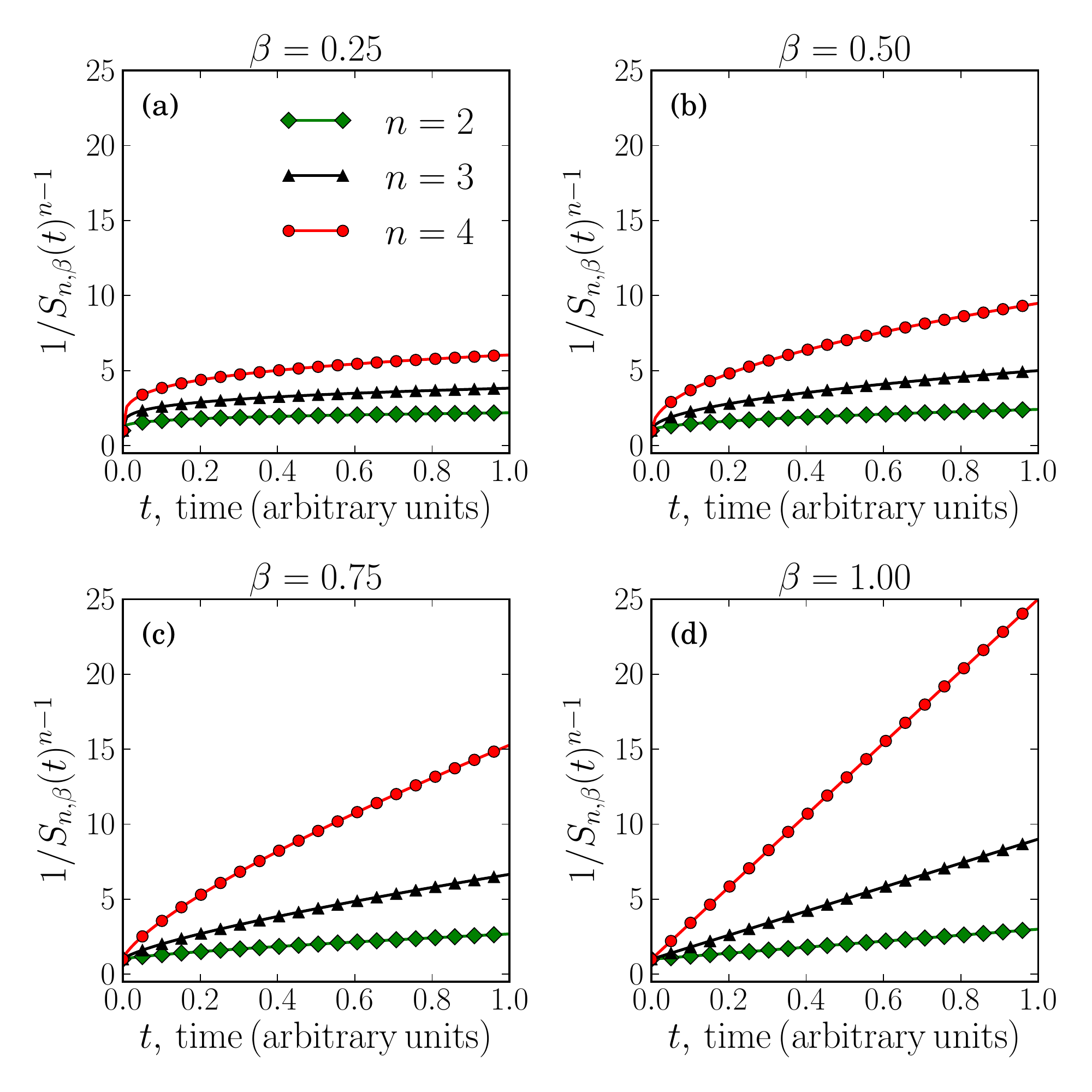}

  \caption{\label{fig:kww-survival}When the disorder parameter, $\beta$, is (a)
$0.25$, (b) $0.50$, or (c) $0.75$ the function $1/S_{n,\beta}^{n-1}(t)$ is
non-linear in time. This non-linearity motivates the use of an effective
time-dependent rate coefficient for the overall decay when the rate equation is
known to be $n^\textrm{th}$-order in the decaying population. Traditional
$n^\textrm{th}$-order kinetics corresponds to the special case of (d) $\beta =
1$, where the rate coefficient is time independent and comes from the slope of
these plots, $\omega_nC_n(0)^{n-1}$. Here $C_n(0) = 2$ and $\omega_n=1$ in
arbitrary units. The effective, time-dependent rate coefficient is proportional
to the slope at a given time.}

\end{figure}
We introduce disorder and a time-dependent rate coefficient into
``higher-order'' kinetics through the $n^\textrm{th}$-order survival function
\begin{equation}
  \label{eqn:sfunc}
  S_{n,\beta}(t) =
  \sqrt[n-1]{\frac{1}{1+(n-1)\left[\omega_n C_n(0)^{n-1}t\right]^\beta}}.
\end{equation}
Again there is a characteristic rate constant $\omega_n$ and disorder parameter
$0<\beta\leq 1$. We chose $\omega_n = 1$ and a fixed observational time scale
to simplify the presentation of our results. Plonka and coworkers interpret the
stretching parameter for $n=2$, as the result of a time-dependent energy
barrier height, and as the result of a superposition of second-order
decays.~\cite{PlonkaBC89} For $n\geq 2$, the characteristic kinetic plot -- a
graph of $1/S_{n,\beta}^{n-1}$ versus $t$ -- is shown in
Figure~\ref{fig:kww-survival}.  As these graphs illustrate, decreasing the
parameter $\beta$ stretches the survival plot for $n^{\textrm{th}}$-order
kinetics, just as it does for first-order processes. Another similarity with
first-order kinetics is that the limit $\beta = 1$ corresponds to traditional
kinetics, the absence of disorder, and when the effective rate coefficient,
$k_n(t)\to \omega_nC_n(0)^{n-1}$, corresponds to the constant slope of the
graph.


\begin{figure}[t]
  \centering
  \includegraphics[width=0.98\columnwidth]{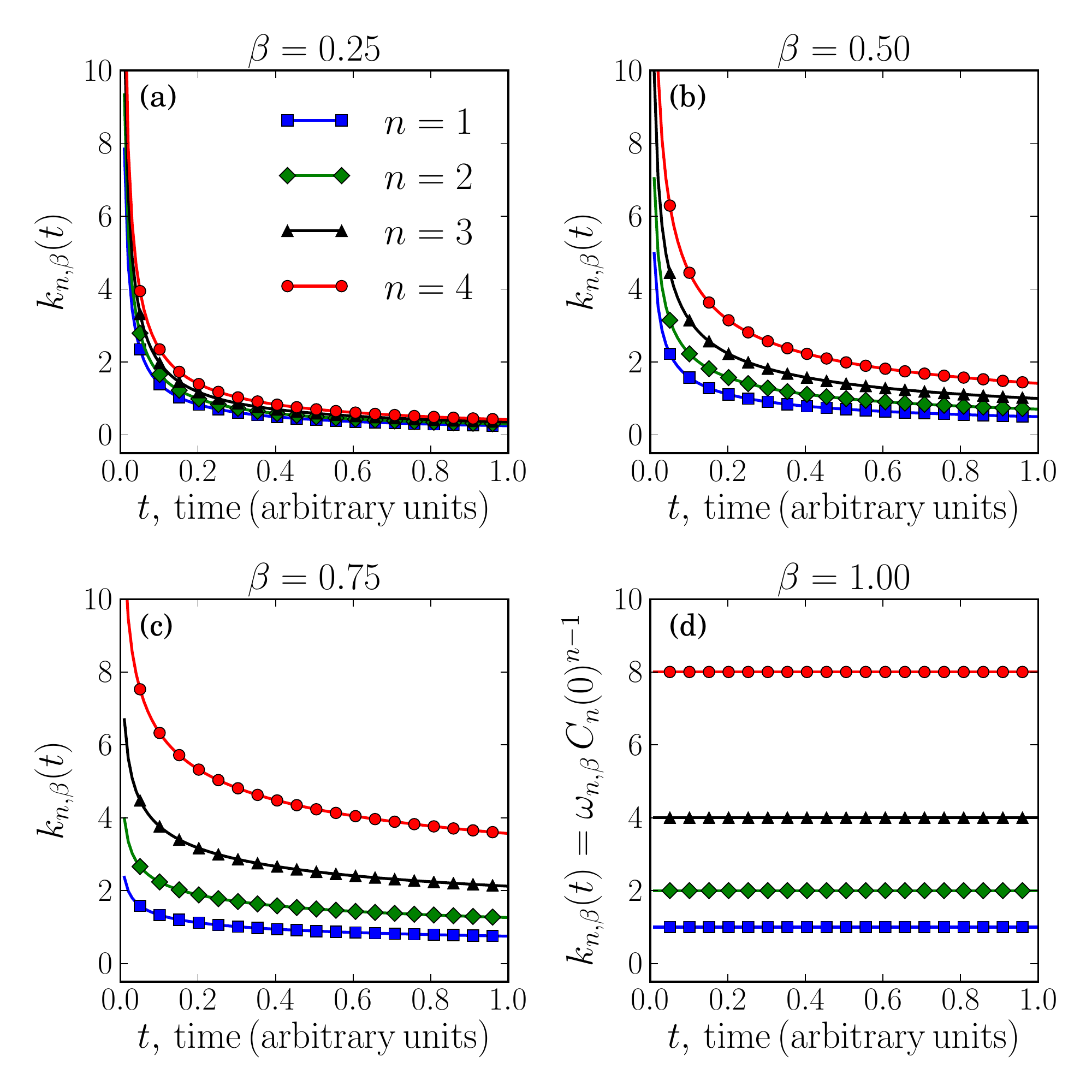}

  \caption{\label{fig:kww-kt}Time-dependence of the effective rate coefficient
$k_{n,\beta}(t)$ for varying reaction orders $n=1$, $2$, $3$, and $4$ when
$\beta$ is (a) $0.25$, (b) $0.50$, (c) $0.75$, and (d) $1.00$. The
$k_{n,\beta}(t)$ of different order $n$ become the same function of time as
$\beta\to 0$ and time independent as $\beta\to 1$. In all cases the initial
concentration is $C_n(0) = 2.0$ and the rate constant is $\omega_n = 1$ in
arbitrary units.}


\end{figure}

At and away from the limit $\beta=1$, the survival functions satisfy rate
equations
\begin{equation}
  \frac{dS_{n,\beta}(t)}{dt} = -k_{n,\beta}(t)S_{n,\beta}(t)^n
\end{equation}
with the same form as Equation~\ref{eqn:disorderedratelaw}. The effective rate
coefficients in these cases are
\begin{eqnarray}
  \label{eqn:kt}
  k_{n,\beta}(t)\nonumber
  &=& \beta\left[\omega_n C_n(0)^{n-1}t\right]^\beta /t\\
  &=& C_n(0)^{\beta\left(n-1\right)}k_{1,\beta}(t).
\end{eqnarray}
This definition of the time-dependent rate coefficients differs from that of
other authors~\cite{Metzler98} to satisfy the rate equations above. 
Keep in mind, we
define the effective rate coefficient in terms of the dimensionless survival
function $S_{n,\beta}(t)$. If we instead define the effective rate coefficient
in terms of the concentration, $C_{n,\beta}(t)$, then the effective rate
coefficient becomes a constant as $\beta\to 1$, with dimensions of
[concentration]$^{-(n-1)}$\,$\cdot$\,[time]$^{-1}$. We will use the former
definition of the effective rate coefficient because the subsequent results
will depend on the initial concentration of the decaying species, which is
consistent with other results in traditional kinetics (e.g., the
half-life depends on the initial concentration in kinetics with $n\geq
2$).~\cite{Houston06}

\subsection{Inequality for disorder in irreversible kinetics}

Having laid out the details and main aspects of the model, we turn to the
application of the theory. For this class of disordered kinetics, it can be
difficult to distinguish the survival curves for different $n$ when $\beta$ is
small. For example, experimental $\beta$ values around $0.1$ are necessary to
fit peptide folding data to a stretched exponential.~\cite{Metzler98,Plonka00}
In this case, the data are also well fit by an asymptotic power law decay (the
integrated form of a second-order rate equation). Our results support this
fact, and show it is also true of irreversible decay for higher $n$.
Figure~\ref{fig:kww-kt}, shows the time dependence of $k_{n,\beta}(t)$ for
$n=1$, $2$, $3$, and $4$ when $\beta$ is (a) $0.25$, (b) $0.50$, (c) $0.75$,
and (d) $1.00$. These data are for an initial concentration of $C_n(0) = 2.0$
and $\omega_n = 1.0$ in arbitrary units. As $\beta\to 0$ the effective rate
coefficient $k_{n,\beta}(t)\to k_\beta(t)$, and the corresponding survival
plots of all reaction order converge, $S_{n,\beta}(t)\to S_{\beta}(t)$. That
is, it can become difficult to distinguish the order of a reaction when $\beta$
is small, which can be taken to mean decay events are highly cooperative or
correlated.

When a $\beta$ approaching zero is necessary to fit the available survival
data, processes with an order $n \geq 2$ may not be distinguishable from
first-order processes. That is, for this widespread type of kinetics,
higher-order kinetics may be an equally valid description of experimental data
when $\beta$ is small, unless the modeller has more information about the
underlying mechanism to suggest a particular reaction order. The greater the
value of $\beta$ is, than say $0.1$, the easier it should be to discern rate
equations with $n \geq 2$ within the experimental uncertainty.  Exactly how
small $\beta$ must be for the reaction order to lose meaning will depend on the
value of $\omega_n$, which we take to be unity, and the observational time
scale. When the rate equations of different orders are distinguishable, we will
show there is an interesting interplay between order, $n$, and the disorder
parameter, $\beta$, that we can disentangle with the measure of dynamical
disorder, $\mathcal{J}_{n,\beta}-\mathcal{L}_{n,\beta}$. To do this, we will
need the length and divergence.

From the effective rate coefficient for $n\geq 2$, the statistical length is
\begin{equation}
  \mathcal{L}_{n,\beta}(\Delta{t}) =
  \left[\omega_nC_n(0)^{n-1} t\right]^{\beta}\big|_{t_i}^{t_f}
\end{equation}
and the divergence is
\begin{multline}
  \frac{\mathcal{J}_{n,\beta}(\Delta{t})}{\Delta t} = 
  \beta^{2}\left[\omega_nC_n(0)^{n-1}\right]^{2\beta}\\\times
  \begin{cases}
    \displaystyle \frac{t^{2\beta-1}}{2\beta-1}\bigg|_{t_i}^{t_f}
    & \text{if } \beta \neq \tfrac{1}{2} \\[10pt]
    \displaystyle \ln t\big|_{t_i}^{t_f}
    & \text{if } \beta = \tfrac{1}{2}.
  \end{cases}
\end{multline}
The effective, time-dependent rate coefficient depends on the initial
concentration, which carries through to $\mathcal{J}_{n,\beta}$ and
$\mathcal{L}_{n,\beta}$. Comparing the inequality for first-order
kinetics~\cite{FlynnZG14} to that of higher-order kinetics, we see the
inequalities are related:
\begin{equation}
  \mathcal{J}_{n,\beta} - \mathcal{L}_{n,\beta}^2 = C_n(0)^{2\beta\left(n-1\right)}
  \left[\mathcal{J}_{1,\beta} - \mathcal{L}_{1,\beta}^2\right].
\end{equation}
The initial concentration, and its dependence upon $n$ and $\beta$,
distinguishes the inequality of higher-order kinetics from first-order
kinetics. We will examine this relation as a means of determining the reaction
order, $n$, in disordered kinetics after exploring the effect of $n$ and
$\beta$ on the inequality.

\subsection{Disorder and the parameters $\boldsymbol{\beta}$ and $\boldsymbol{n}$}

\begin{figure}[!t]
  \centering
  \includegraphics[width=0.98\columnwidth]{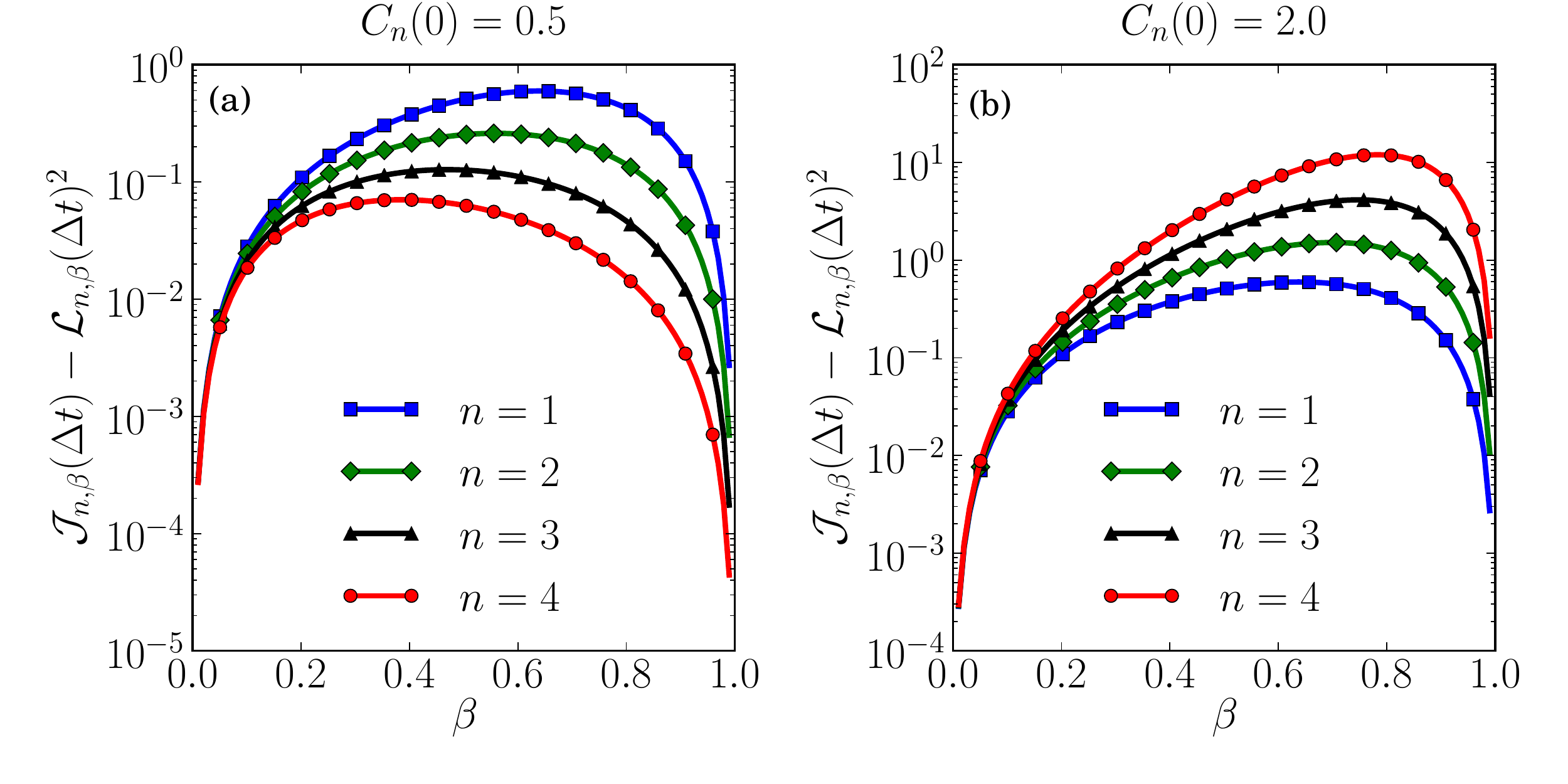}

  \caption{\label{fig:kww-vary-beta}The magnitude of the inequality,
$\mathcal{J}_{n,\beta}-\mathcal{L}^2_{n,\beta}$, on a logarithmic scale over
the range of the disorder parameter, $\beta$, when the initial concentration
$C_n(0)$ is (a) $0.5$ and (b) $2.0$. Here $\omega_n=1$, $t_i=1$, and $t_f=10$
in arbitrary, but consistent, units. Increasing the reaction order, $n$, can
increase or decrease the dynamical disorder, as measured by the inequality,
depending on the initial concentration. The inequality correctly reflects the
absence of dynamical disorder in the limits $\beta\to 0$ and $\beta = 1$.}

\end{figure}

From these results we can show $\mathcal{J}_{n,\beta}-\mathcal{L}_{n,\beta}^2$
measures dynamical disorder for irreversible decay kinetics with $n\geq 1$. A
minimum criterion a reasonable measure of disorder must have is that it be zero
in the absence of disorder. For the present model, this means the measure must
then be zero when $\beta = 1$ and tend to zero as $\beta\to 0$, since the rate
coefficient $k_{n,\beta}(t)$ is time independent in these limits. Indeed, we
find the inequality satisfies this requirement, and the lower bound
$\mathcal{J}_{n,\beta}-\mathcal{L}_{n,\beta}^2=0$ holds when $\beta = 0$, $1$
irrespective of the reaction-order, $n$. It is important to note that this is
true when $k_n(t)$ is defined as in Equation~\ref{eqn:kt}.

Away from the limits $\beta = 0$ and $1$, we find the statistical length and
divergence are not equal, $\mathcal{J}_{n,\beta}-\mathcal{L}_{n,\beta}^2\geq 0$
(Figure~\ref{fig:kww-vary-beta}). Interestingly, for all $n$, their difference
initially increases with decreasing $\beta$ from one to zero, since this
stretches the survival plot and increases the variation in the effective rate
coefficient over the observational time window. However, as $\beta$ decreases
further, there is eventually a point at which the rate coefficient is
effectively constant over the majority of the observational time window. This
leads to an eventual decrease in disorder when stretching
$n^{\textrm{th}}$-order kinetics due to the separation of decay into ``fast''
and ``slow'' components. The inequality accurately reflects this effect and,
consequently, has a maximum in its dependence upon $\beta$, tending to the
lower bound $\mathcal{J}_{n,\beta}-\mathcal{L}_{n,\beta}^2=0$ as $\beta\to 0$.
Another interpretation of the inequality is as a measure of the deviations of
the $n^{\textrm{th}}$-order characteristic kinetic plot from linear during
observational time window, and the portion of the time window when those
deviations occur. The maximum in the inequality results from the competition of
these factors. These findings are evidence that the inequality measures the
amount of dynamical disorder in irreversible decay processes of any
reaction-order.


Figure~\ref{fig:kww-vary-beta} shows the dependence of the degree of dynamical
disorder on the disorder parameter, $\beta$. The inequality reveals a main
feature of dynamical disorder: the disorder parameter, $\beta$ does not
linearly increase the amount of dynamical disorder in irreversible relaxations
describable by stretched exponential and power law ($n\geq 2$,
Equation~\ref{eqn:sfunc}) decay. Further, the amount of dynamical disorder
measured by the inequality does not uniquely determine the value of $\beta$.
Increasing (for $\beta$ values greater than that corresponding to the maximum
in $\mathcal{J}_{n,\beta}-\mathcal{L}^2_{n,\beta}$) or decreasing $\beta$ (for
$\beta$ values less than that corresponding to the maximum in
$\mathcal{J}_{n,\beta}-\mathcal{L}^2_{n,\beta}$) can improve the approximation
of the decay over the observational time scale with a single rate coefficient.
We take this as further evidence that the inequality is a useful measure of
dynamical disorder.

Comparing panel (a) and (b) in Figure~\ref{fig:kww-vary-beta}, we see the
initial concentration dictates whether increasing the reaction-order increases
or decreases the inequality, and so, the dynamical disorder for a given
$\beta$, $\omega_n$, and observational time scale. Panel (a) shows that for an
initial concentration $C_n(0)<1$,
$\mathcal{J}_{n,\beta}-\mathcal{L}_{n,\beta}^2$ decreases with reaction order,
$n$, at a given $\beta\neq 0$ or $1$. This suggests that for a given $\beta$,
higher-order the kinetics will have an effective rate coefficient
$k_{n,\beta}(t)$ that varies less over the observational time scale: there is
less dynamical disorder with increasing reaction order. In contrast, panel (b)
shows the converse is true when the initial concentration $C_n(0)>1$:
$\mathcal{J}_{n,\beta}-\mathcal{L}_{n,\beta}^2$ increases with reaction order,
$n$, suggesting $k_{n,\beta}(t)$ varies more over the observational time scale
the higher-order the kinetics. Overall, higher reaction-order processes can
have more or less disorder for a given ($\beta$, $\omega_n$, $\Delta t$)
depending on the initial concentration. The reaction-order is not the opposite
of dynamical disorder.

Among the growing number of disordered systems,~\cite{Plonka01,Ross08} a couple
of interesting applications of our approach would be peptide
folding~\cite{Metzler98} and low-temperature recombination reactions, such as
the recombination of carbon monoxide with myoglobin,~\cite{AustinBEFG75} which
have small stretching parameters ($\beta \approx 0.1$). Inspecting
Figure~\ref{fig:kww-vary-beta} around $\beta = 0.1$ we find there are
$n$-dependent differences in the
$\mathcal{J}_{n,\beta}-\mathcal{L}_{n,\beta}^2$ curves. At $\beta=0.1$ the
inequality is small and only weakly dependent on the reaction order. This is
quantitative evidence of why experimental peptide folding data that is well fit
by a stretched exponential ($n = 1$) with a $\beta$ value around $0.1$ is also
fit by an asymptotic power law decay ($n = 2$). The near equivalence of
$\mathcal{J}_{1,\beta}-\mathcal{L}^2_{1,\beta}$ and
$\mathcal{J}_{2,\beta}-\mathcal{L}^2_{2,\beta}$ for reactions of different $n$
suggests the degree of disorder is approximately the same within experimental
uncertainties. This result also suggests higher-reaction-orders may fit this
data.

\begin{figure}[!t]
  \centering
  \includegraphics[width=0.98\columnwidth]{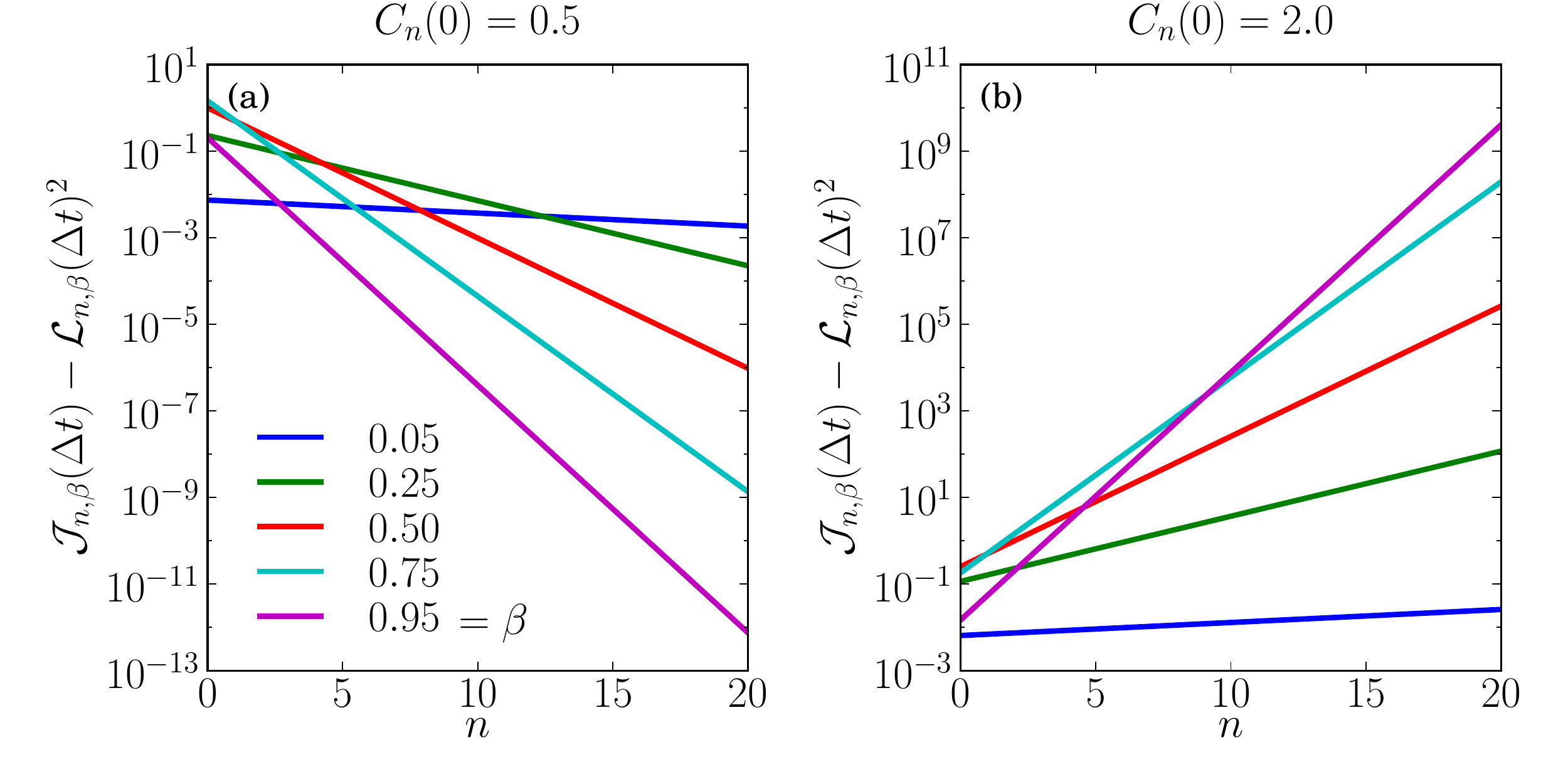}

  \caption{\label{fig:kww-vary-n}The magnitude of the inequality,
$\mathcal{J}_{n,\beta}-\mathcal{L}^2_{n,\beta}$, on a logarithmic scale over
the range of the reaction-order, $n$, when the initial concentration $C_n(0)$
is (a) $0.5$ and (b) $2.0$. Here $\omega_n=1$, $t_i=1$, and $t_f=10$ in
arbitrary, but consistent, units. The disorder parameter, $\beta$, controls how
strongly the dynamical disorder, as measured by the inequality, depends on the
reaction-order, $n$, regardless of the initial concentration.  When $\beta =
1$, $\mathcal{J}_{n,\beta}=\mathcal{L}^2_{n,\beta}$ for all $n$.}

\end{figure}

Using this model for irreversible decay, we can also consider  the dependence
of the inequality on the reaction-order parameter, $n$.
Figure~\ref{fig:kww-vary-n} shows that $\beta$ determines how strongly the
inequality (dynamical disorder) depends on the reaction order. We see
increasing $\beta$ strengthens the dependence of the dynamical disorder on the
reaction-order -- whether the dynamical disorder increases or decreases with
$n$. In particular, as $\beta \to 0$ the degree of dynamical disorder, as
measured by the inequality, becomes more uniform over the range of $n$; the
dynamical disorder becomes independent of the reaction order. Furthermore, this
finding is independent of the initial concentration, as the two panels show. We
also saw that as $\beta\to 0$, the $n$-dependent variations in the effective
rate coefficient vanish, $k_{n,\beta}(t)\to 0$ in Figure~\ref{fig:kww-kt},
which is why $\mathcal{J}_{n,\beta}-\mathcal{L}_{n,\beta}^2$ also becomes
independent of $n$. Since the parameter $\beta$ determines the rate of change
of $\mathcal{J}_{n,\beta}-\mathcal{L}_{n,\beta}^2$ with $n$, for small $\beta$,
the lack of variation of the inequality with $n$ suggests a fundamental
difficulty in determining the reaction order. These results are another
reflection of the difficulty in determining the reaction order in
non-exponential kinetics when $\beta$ is small.

\subsection{Disorder and the initial concentration, $\boldsymbol{C_n(0)}$}

\begin{figure}[!t]
  \centering
  \includegraphics[width=0.98\columnwidth]{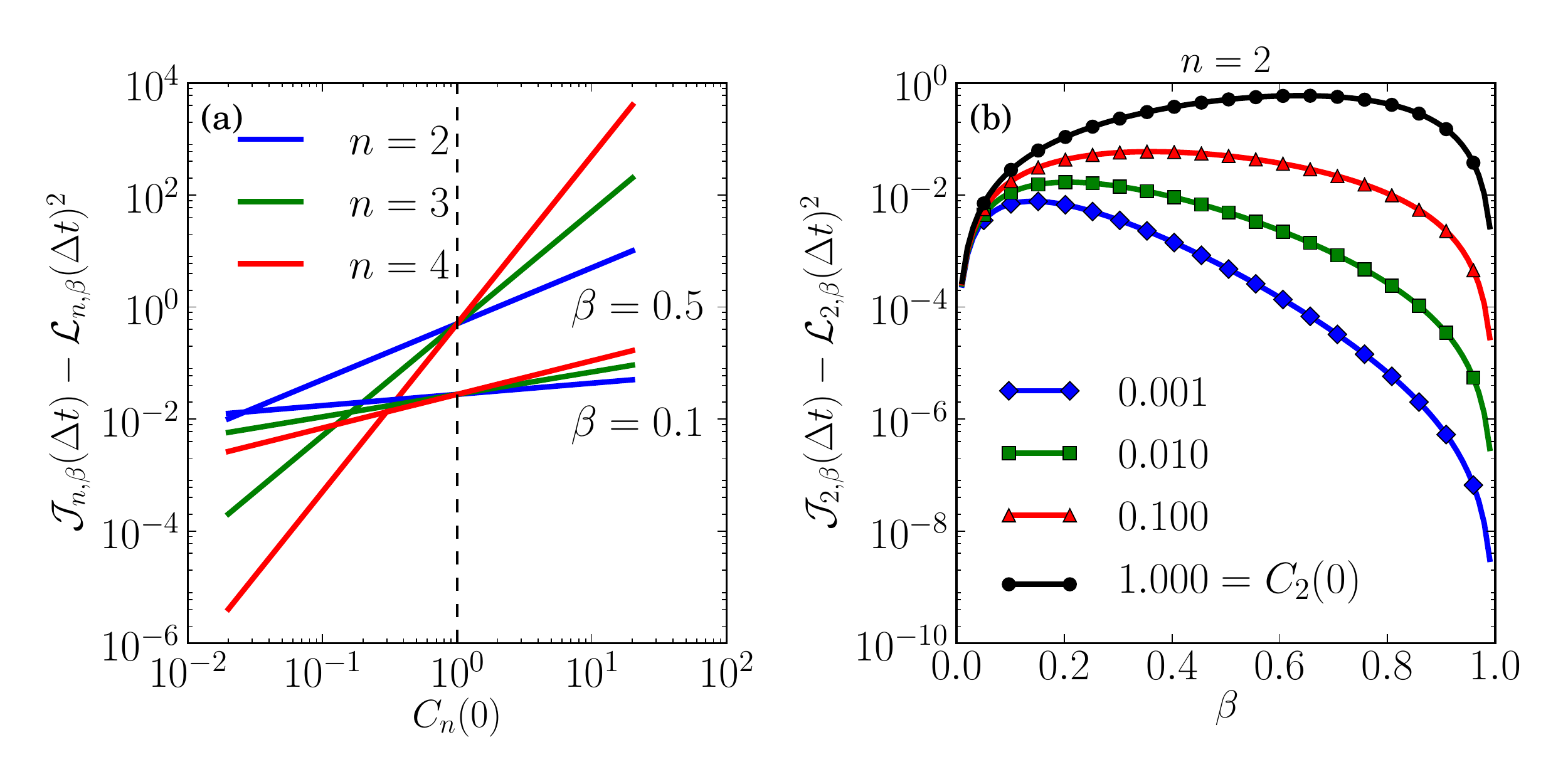}

  \caption{\label{fig:kww-inequality-final}(a) A log-log plot of the
inequality, $\mathcal{J}_{n,\beta}-\mathcal{L}^2_{n,\beta}$, as a function of
the initial concentration, $C_n(0)$, for $n=2-4$ and $\beta=0.1$, $0.5$. (b)
The inequality, $\mathcal{J}_{2,\beta}-\mathcal{L}^2_{2,\beta}$, for
second-order kinetics on a log scale versus the disorder parameter, $\beta$,
for initial concentrations, $C_n(0)$, spanning four orders of magnitude and a
fixed observational time scale ($t_i=1$, $t_f=10$).}

\end{figure}

In traditional macroscopic kinetics, initial rate methods are a standard means
of determining the order(s) in the rate equation. Initial rate methods involve
varying the initial concentration of a particular reactant, and measuring the
initial rate. Through a series of such measurements, an experimenter can find
the order of a hypothesized rate equation through algebraic manipulation, and
gain clues about the mechanism of the reaction.  We take a similar approach for
the present model of disordered kinetics.
Figure~\ref{fig:kww-inequality-final}(a) shows a log-log plot of
$\mathcal{J}_{n,\beta}-\mathcal{L}^2_{n,\beta}$ as a function of the initial
concentration, $C_n(0)$, for $n=2-4$ and $\beta=0.1$, $0.5$. The inequality is
over the observational time window, $t_i = 1$ and $t_f = 10$.

Most immediate from these data is that the magnitude of the inequality, and the
dynamical disorder it represents, directly correlates with $C_n(0)$ for all
$\beta$ and $n$. At a given $\beta$,
$\mathcal{J}_{n,\beta}-\mathcal{L}_{n,\beta}^2$ may increase or decrease with
reaction-order, $n$, depending on the initial concentration: the inequality
increases with $n$ for $C_n(0)>1$ and decreases for $C_n(0)<1$, as shown in
Figures~\ref{fig:kww-vary-beta} and~\ref{fig:kww-vary-n}. Also of note is that
while $\mathcal{J}_{n,\beta}-\mathcal{L}_{n,\beta}^2$ increases more rapidly
for $n=4$ than $n=2$, the difference can be offset by a small $\beta$, e.g.,
$0.1$. This detail reflects the more general finding that the ``strength'' of
the increase or decrease of the inequality depends on $\beta$; decreasing the
value of $\beta$ suppresses the concentration dependence of the dynamical
disorder (inequality) for all reaction orders, and the differences in the
inequality for different $n$. Thus, even the initial concentration dependence
may not be sufficient to distinguish the disordered kinetics of reactions with
different reaction order.

Figure~\ref{fig:kww-inequality-final}(b), shows how the inequality for
second-order kinetics, $\mathcal{J}_{2,\beta}-\mathcal{L}^2_{2,\beta}$, varies
with the disorder parameter, $\beta$, for initial concentrations,
$\omega_nC_n(0)$, spanning four orders of magnitude and a fixed observational
time scale. The maximum in the inequality at $\beta_\textrm{max}$ shifts to
larger $\beta$ values with increasing $C_2(0)$. Similar behavior results from
increasing $\omega_2$, which we have set to one. For
$\beta<\beta_\textrm{max}$, $\beta$ and $n$ have the opposite effect on the
dynamical disorder: increasing $\beta$ and decreasing $n$ will increase the
dynamical disorder. But if $\beta>\beta_\textrm{max}$, $\beta$ and $n$ have the
same effect on the dynamical disorder: decreasing $\beta$ and decreasing $n$
will increase the dynamical disorder. We conclude that $n$ and $\beta$ are not
strictly opposed parameters of order and disorder, and the inequality, a
measure of dynamical disorder is necessary to disentangle these subtleties of
disordered kinetics.

Measuring the initial concentration dependence of the inequality is a
straightforward procedure to try to determine the reaction order, and a clue to
the underlying mechanism, even if the kinetics is disordered. However, the
inequality reveals that in disordered kinetics there is a fundamental
difficulty in identifying a unique assignment of the reaction order from
survival data.  The difficulty will surely increase when there is uncertainty
in the survival function, as will be the case in experiments and simulations.
Under these circumstances, the fluctuations in the survival function will also
likely impact the inequality, though here the concentrations and survival
functions are non-fluctuating quantities. Even under ideal conditions, as we
have shown, the reaction order is less meaningful when a small disorder
parameter, $\beta$, is necessary. 

\section{Conclusions}

In summary, we have considered the inequality,
$\mathcal{J}_n\geq\mathcal{L}_n^2$, between two functions of the effective
time-dependent rate coefficient, the statistical length and divergence. We have
extended the scope of the inequality to irreversible decay processes with any
reaction-order, by appropriately defining $k_n(t)$. A redefinition of $k_n(t)$
preserves the advantageous features of the inequality. First, the inequality
quantifies both the variation of $k_n(t)$ in time and the fraction of the
observation time where those changes occur -- the amount of dynamical disorder.
Second, the lower bound of the inequality, $\mathcal{J}_n = \mathcal{L}_n^2$,
is a condition for traditional kinetics to hold, and for rate coefficient to be
unique. Leveraging this measure, we have found for stretched exponential and
power law decay that the dynamical disorder can increase or decrease with the
reaction order. Higher reaction order does not imply lower dynamical disorder,
and vice-versa. Instead, the parameter $\beta$ determines the ability to
distinguish kinetics of different reaction order in the types of decay
considered. As in traditional kinetics, the initial concentration may be used
to distinguish reactions of different overall order. Here, for dynamically
disordered decay, the initial concentration dependence of the inequality proved
to be a useful way to determine the reaction order (when such a parameter is
meaningful), establish a rate equation, and gain clues about the reaction
mechanism. From our results, it seems justifiable to consider higher-order,
$n>1$, processes for experimental measurements of decay, even when there is
disorder.

\section{Acknowledgements}

This material is based upon work supported by the U.S. Army Research Laboratory
and the U.S. Army Research Office under grant number W911NF-14-1-0359.

%

\end{document}